\documentclass{appolb}
\usepackage{myphi}

\graphicspath{{picts/}}

\let\OLDthebibliography\thebibliography
\renewcommand\thebibliography[1]{
  \OLDthebibliography{#1}
  \setlength{\parskip}{0pt}
  \setlength{\itemsep}{0.6ex plus 0.3ex}
}

\NewMathSymbol{\xF}{x_\text{F}}

\newcommand{\Figref}[1]{Figure~\ref{#1}}
\newcommand{\figref}[1]{Fig.~\ref{#1}}

\begin{document}

\title{Highlights from the \NASixtyOne experiment
  \thanks{Presented at Quark Matter 2022, 5 April 2022, Kraków, Poland}%
}
\author{Antoni Marcinek for the \NASixtyOne Collaboration
  \address{Institute of Nuclear Physics, Polish Academy of Sciences, Kraków, Poland}
}

\maketitle

\begin{abstract}
The \NASixtyOne experiment is a fixed-target, broad acceptance facility at the
CERN SPS. This contribution summarizes the most recent results from the strong
interactions \NASixtyOne programme and presents news on the detector
upgrade in preparation for the future data taking.
The strong interactions programme consists in a two-dimensional scan in
beam momentum (from 13$A$ to 150$A$/158$A$ GeV/$c$, \sNN from 5.1 to 17.3 GeV)
and system size (\pp, \BeBe, \ArSc, \XeLa reactions). The experiment
searches for the second-order critical end-point in the temperature versus
baryo-chemical potential phase diagram and studies the properties of the
onset of deconfinement discovered by its predecessor, NA49 at the CERN SPS.
The presented results include $K/\pi$ multiplicity ratios as a function of
energy and system size,
singly and multi-strange hadron production in \pp reactions, multiplicity
and net-charge fluctuations measured by higher order moments in \pp, \BeBe
and \ArSc collisions, proton and charged hadron intermittency in \ArSc and
\PbPb reactions, HBT measurements in \ArSc
and collective electromagnetic effects in \ArSc collisions.
\end{abstract}

\begin{figure}[htb]
  \centerline{
    \includegraphics[width=0.5\textwidth]{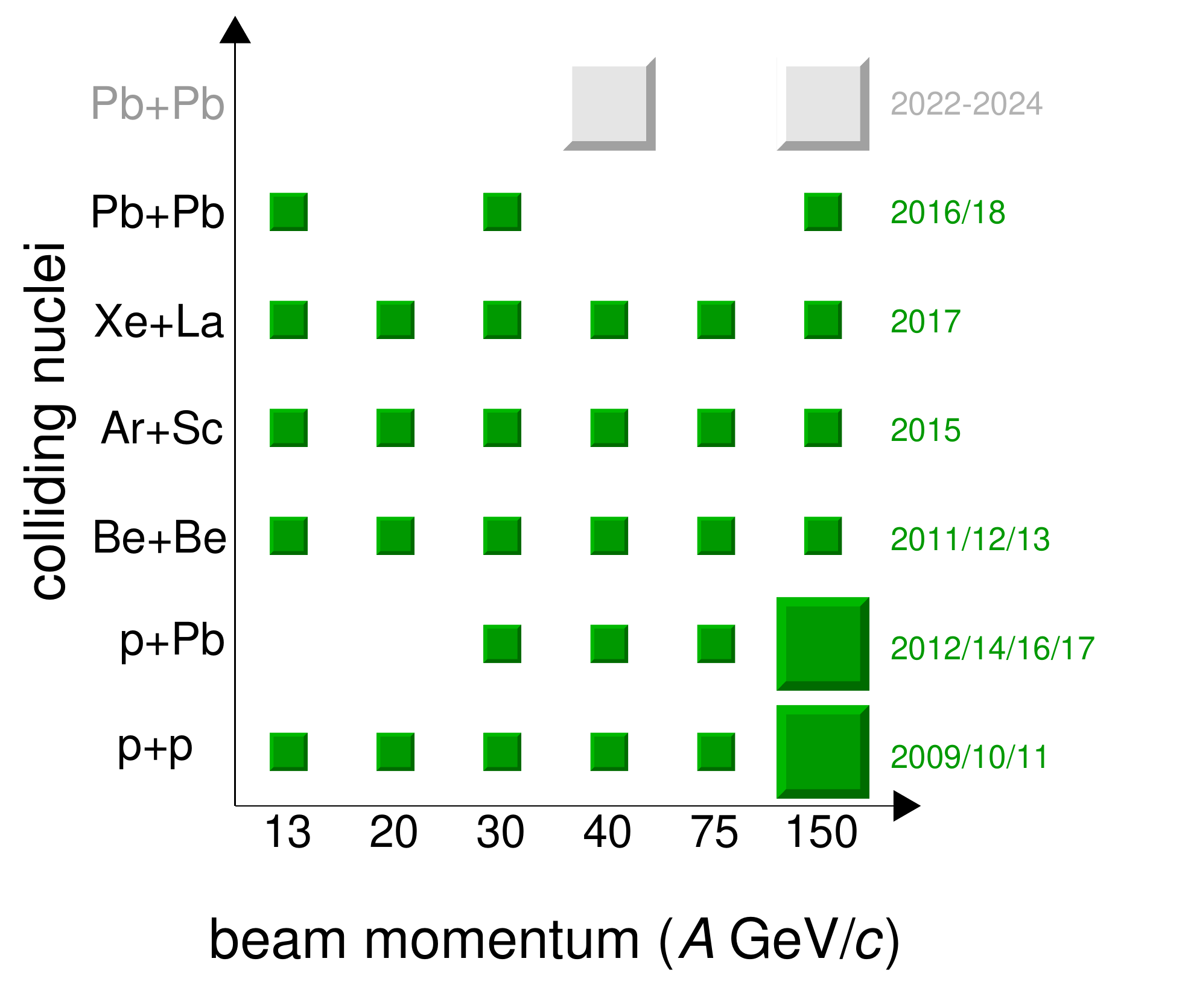}
    \hfill
    \includegraphics[clip, trim=0 50 0 0, width=0.4\textwidth]{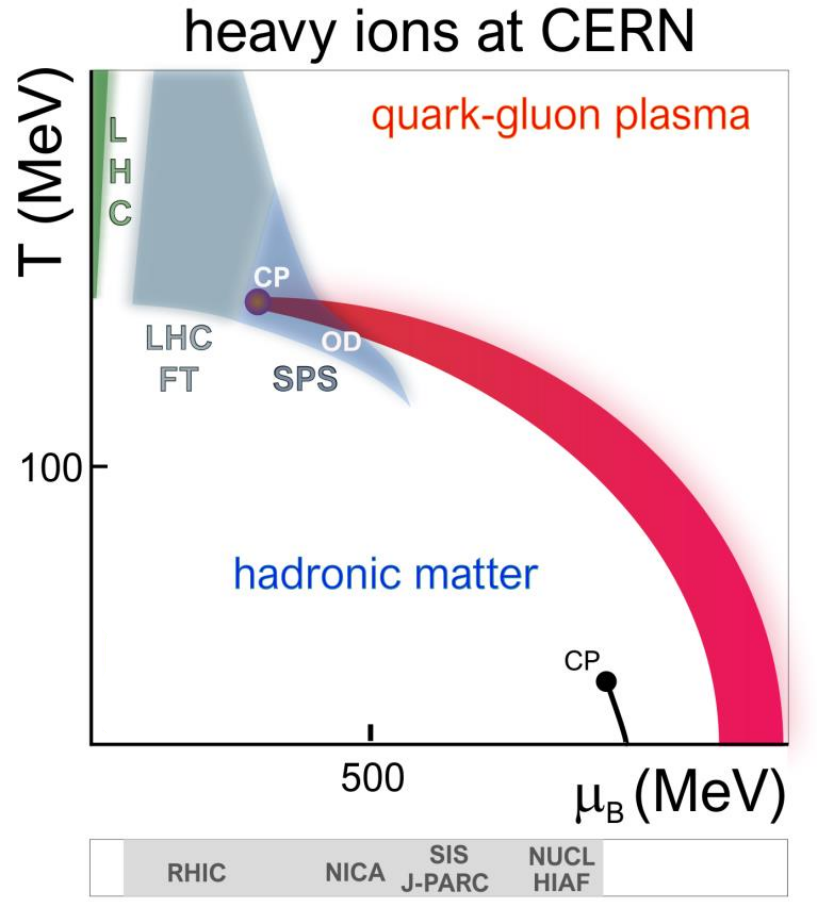}
  }
  \caption{\NASixtyOne system size and energy scan performed within the strong
    interactions programme (left) and the phase diagram of the strongly
    interacting matter (right) studied in this scan.}
  \label{fig:programme}
\end{figure}
\begin{figure}[htb]
  \centerline{%
    \includegraphics[width=0.45\textwidth]{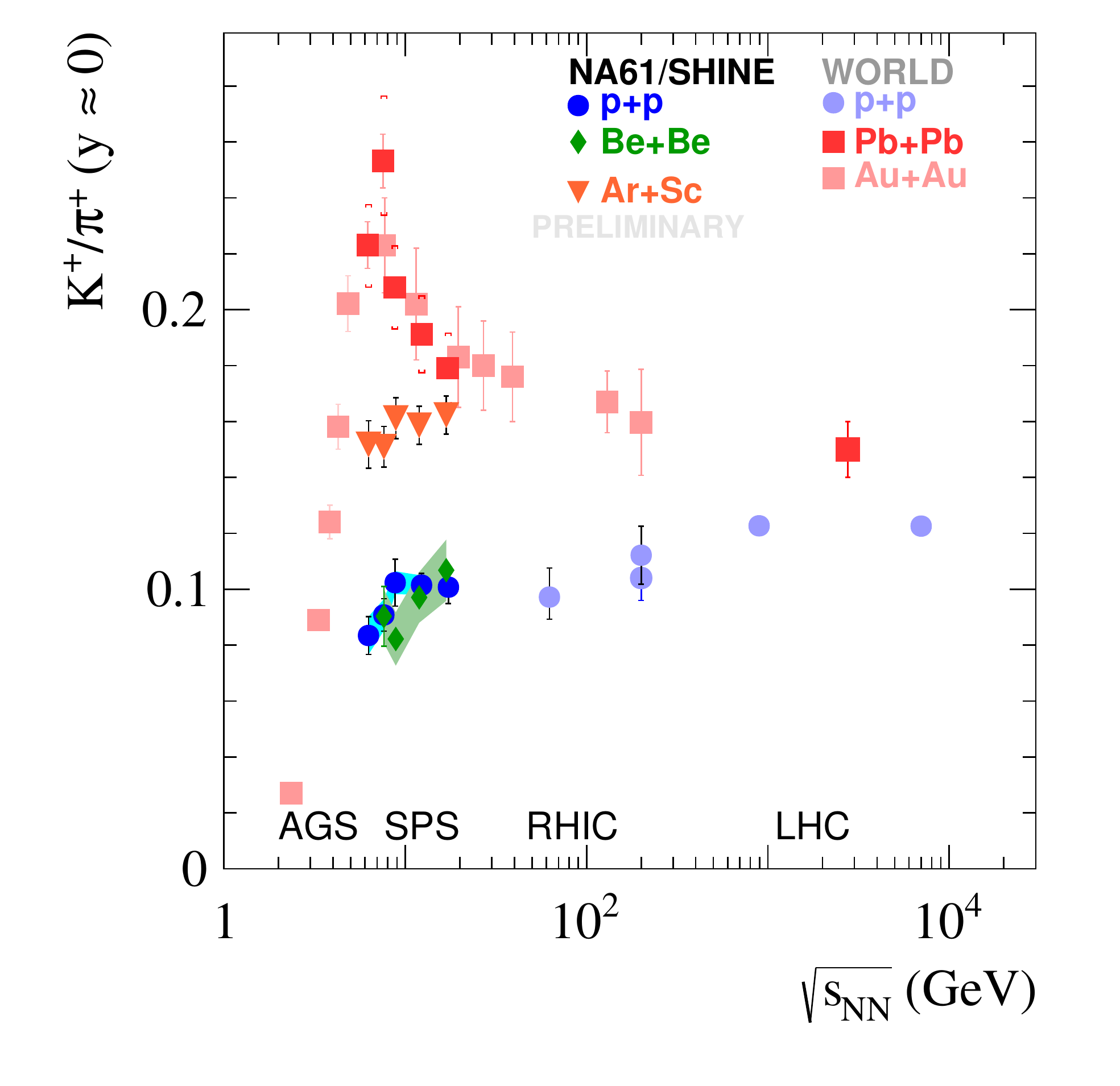}
  }
  \caption[]{System size and energy dependence of the $\Kp/\pip$ ratio in
    midrapidity~\capcite{Maciek} including published, new \NASixtyOne results for
    \BeBe~\capcite{NA61SHINE:2020czq} and preliminary \ArSc collisions.}
  \label{fig:horn}
\end{figure}
\section{Introduction}
The \NASixtyOne experiment at the CERN SPS owes its name
(SPS Heavy Ion and Neutrino Experiment) to a two-fold experimental programme.
On the one hand it conducts precise spectra measurements necessary for cosmic
ray and neutrino experiments. On the other, it investigates the onset of
deconfinement (OD) discovered by its predecessor, the NA49 experiment at the CERN
SPS~\cite{Alt:2007aa} and searches for the critical point (CP) of the strongly
interacting matter.
\par
The latter part of the programme, the strong interactions programme, is the
subject of the present paper. Within it, \NASixtyOne performed a
two-dimensional scan of system size and collision energy
(\figref{fig:programme} left). This corresponds to scanning the phase diagram
of the strongly interacting matter~\cite{Becattini:2005xt}
(\figref{fig:programme} right). It should be noted that due to the search for
CP, which involves fluctuation variables, it is paramount to minimize system
volume fluctuations and consequently the experiment focuses mainly on central
collisions. Hence the system size scan as opposed to the more usual centrality
scan.

\section{Identified hadron spectra in \BeBe and \ArSc collisions}
The published, new results on identified hadron spectra in \BeBe
collisions~\cite{NA61SHINE:2020czq} and preliminary results on identified hadron spectra in \ArSc collisions
are discussed in detail in Ref.~\cite{Maciek}. \Figref{fig:horn} shows the
summary of these studies, the system size and energy dependence of the
$\Kp/\pip$ ratio in midrapidity. A pronounced \enquote{horn} in \PbPb/\AuAu
collisions, associated with the OD~\cite{Gazdzicki:1998vd}, is clearly visible.
There are, however, some unexpected features in this dependence.
First, \pp collisions do not follow a smooth increase with energy, but rather
an abrupt change of the trend at SPS energies.
Second, \BeBe collisions feature similar $\Kp/\pip$ ratio values to \pp, while
\ArSc collisions have no horn structure, but at the top SPS energy converge to
similar (large) ratio values as the \PbPb collisions.
These two observations can be attributed~\cite{Andronov:2022cna} respectively
to two transitions between domains of dominance of specific hadron
production mechanisms: from \emph{hadron resonances} to \emph{strings} when changing collision energy for small systems and from
\emph{strings} to \emph{QGP fireball} (onset of fireball) when changing system size at large energies.
These new \NASixtyOne data allow for the first time to sketch such a diagram of high-energy nuclear collisions 
and emphasize how important the SPS energy range is.

\begin{figure}[htb]
  \centerline{
    \includegraphics[width=0.45\textwidth]{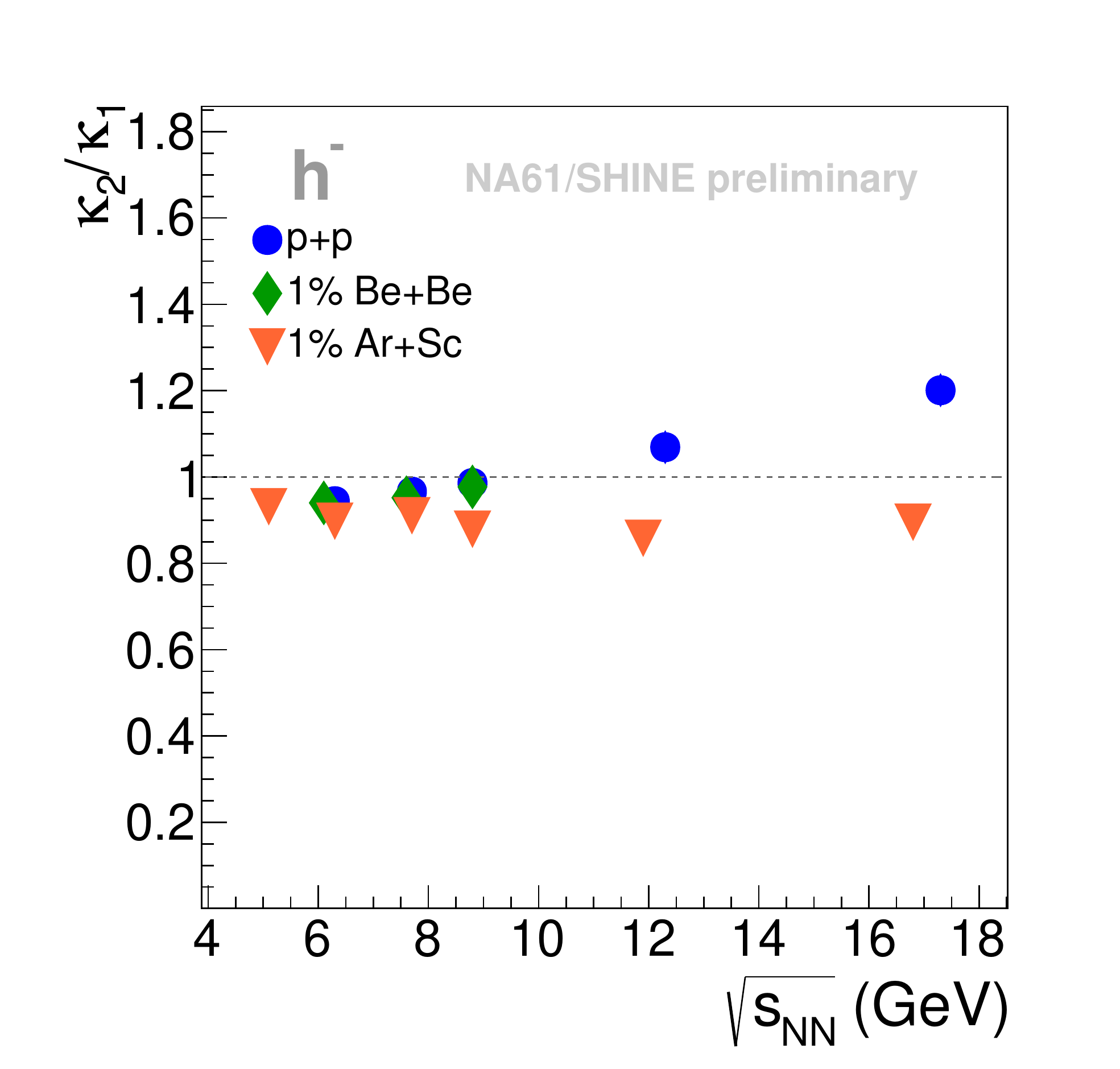}
    \includegraphics[width=0.45\textwidth]{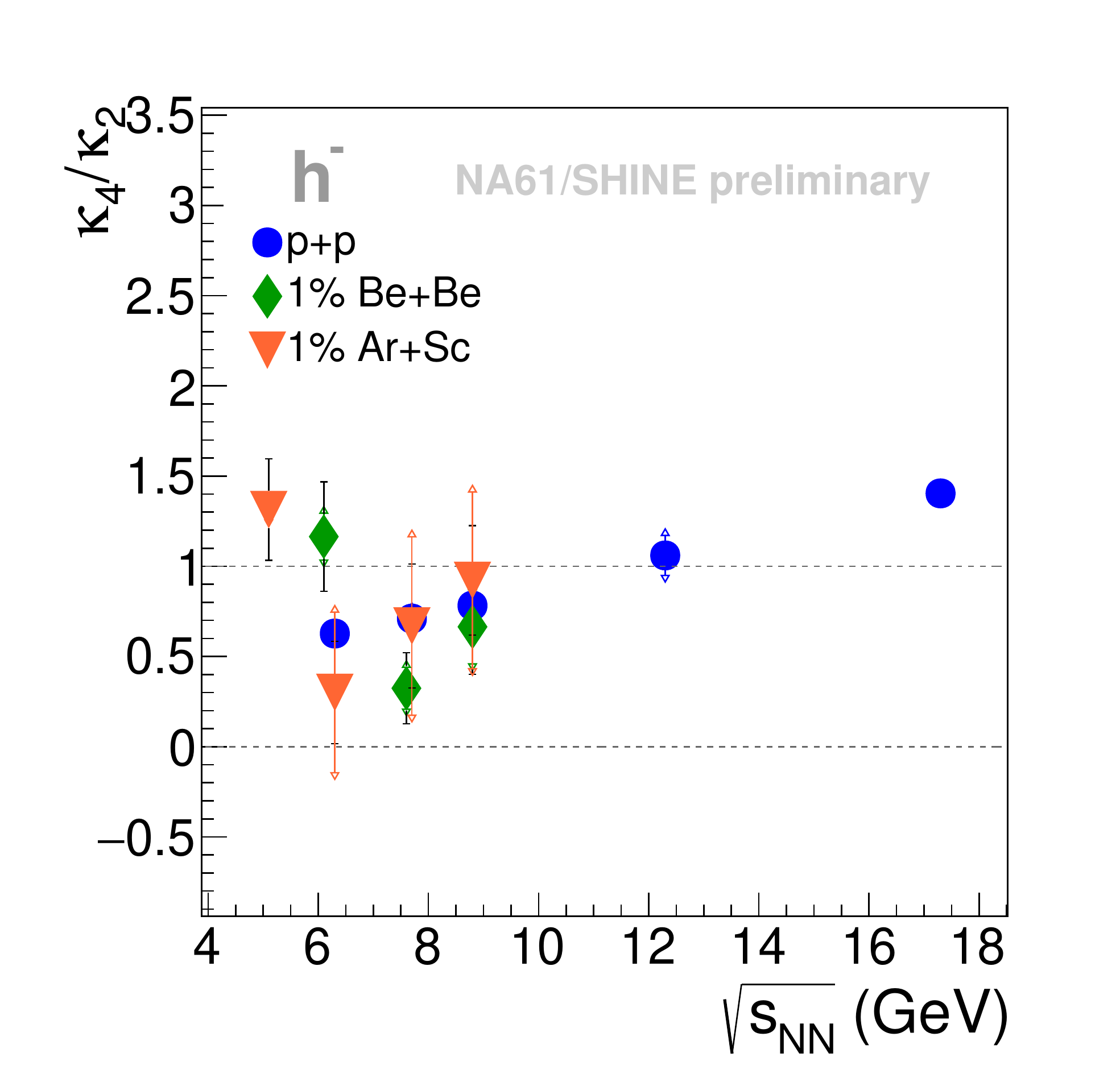}
  }
  \caption{Preliminary results on the system size and energy dependence of the
  multiplicity fluctuations for negatively charged hadrons.}
  \label{fig:mult-fluctuations}
  \centerline{
    \includegraphics[width=0.45\textwidth]{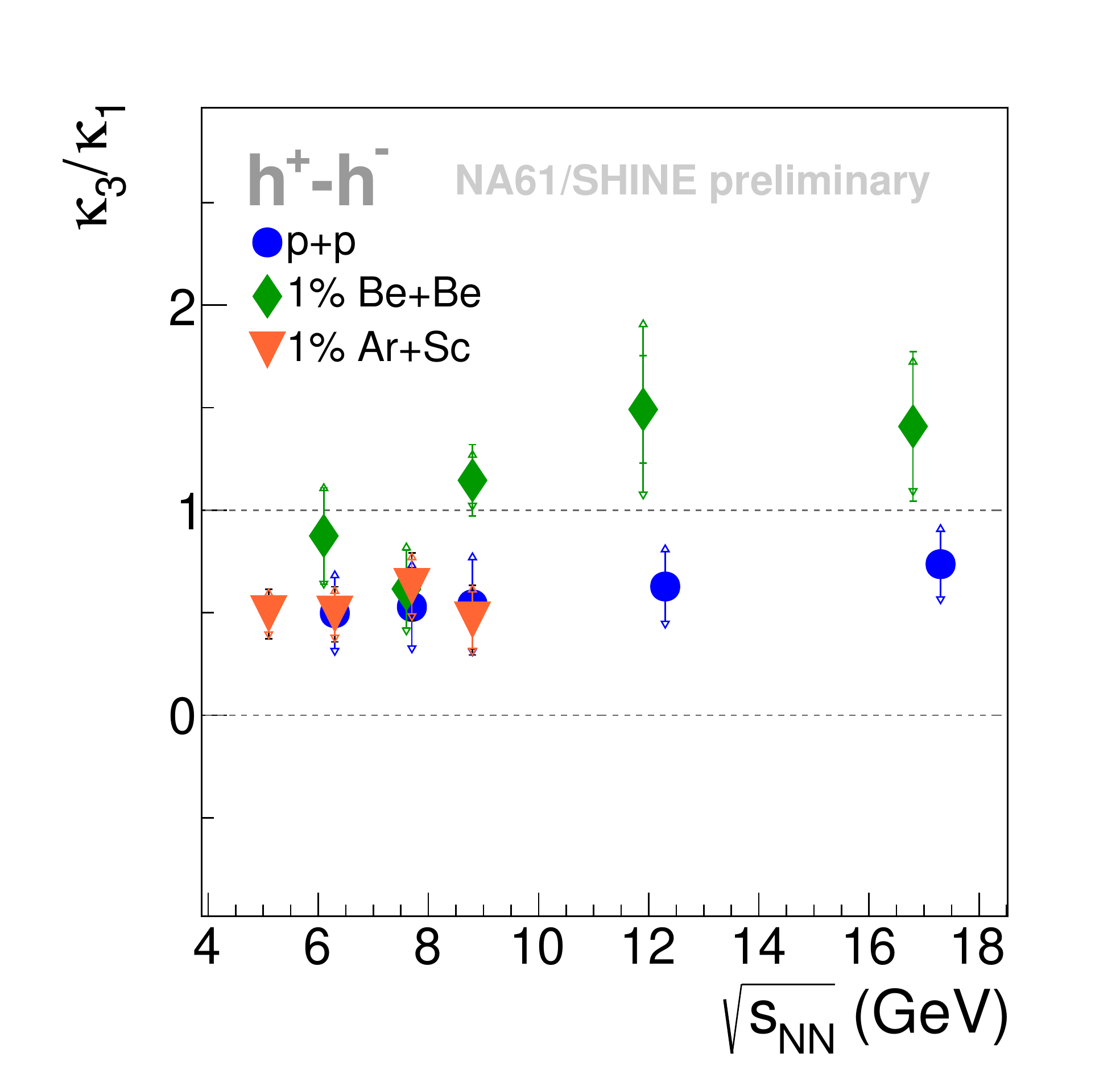}
    \includegraphics[width=0.45\textwidth]{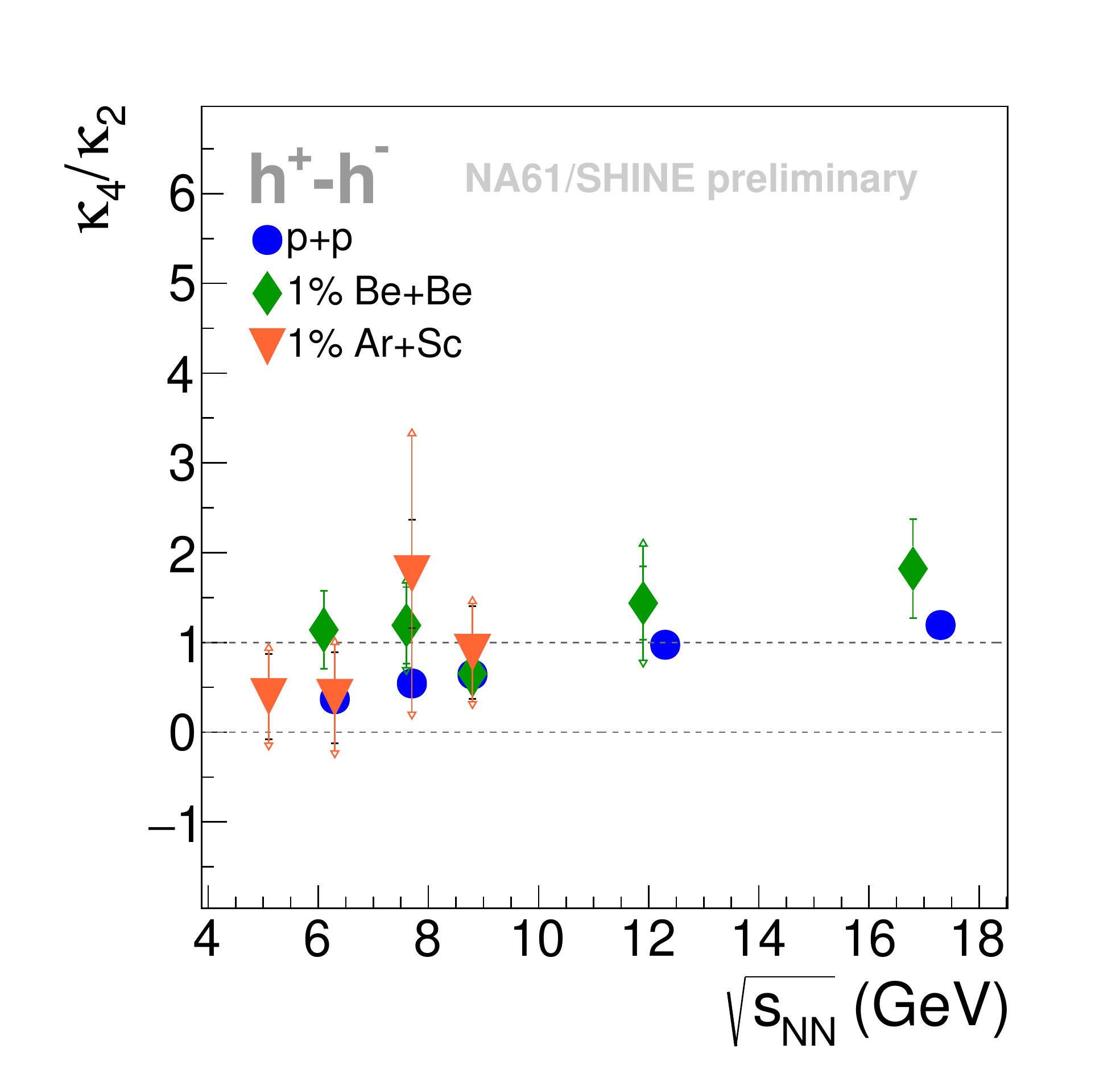}
  }
  \caption{Preliminary results on the system size and energy dependence of the
  net-charge fluctuations for charged hadrons.}
  \label{fig:charge-fluctuations}
\end{figure}
\section{Search for the critical point}
Preliminary results on multiplicity fluctuations of charged hadrons and
net-charge fluctuations of charged hadrons, measured as the ratios of cumulants~\cite{Cumulant}
of the order up to 4, are shown respectively in \figref{fig:mult-fluctuations}
and \figref{fig:charge-fluctuations}. No structure indicating the CP is
visible. While $\kappa_4/\kappa_2$ ratio is consistent for all measured
systems, interesting system size dependence is observed for multiplicity
$\kappa_2/\kappa_1$ and net-charge $\kappa_3/\kappa_1$ ratios. In the first case there
is an increasing difference with collision energy between small systems (\pp
and \BeBe) and the \ArSc system. In the second case there is a difference
between \BeBe and other systems, again increasing with \sNN. Work is still ongoing
to get more data points for \BeBe and \ArSc systems.

\begin{figure}[htb]
  \centerline{
    \includegraphics[width=0.42\textwidth]{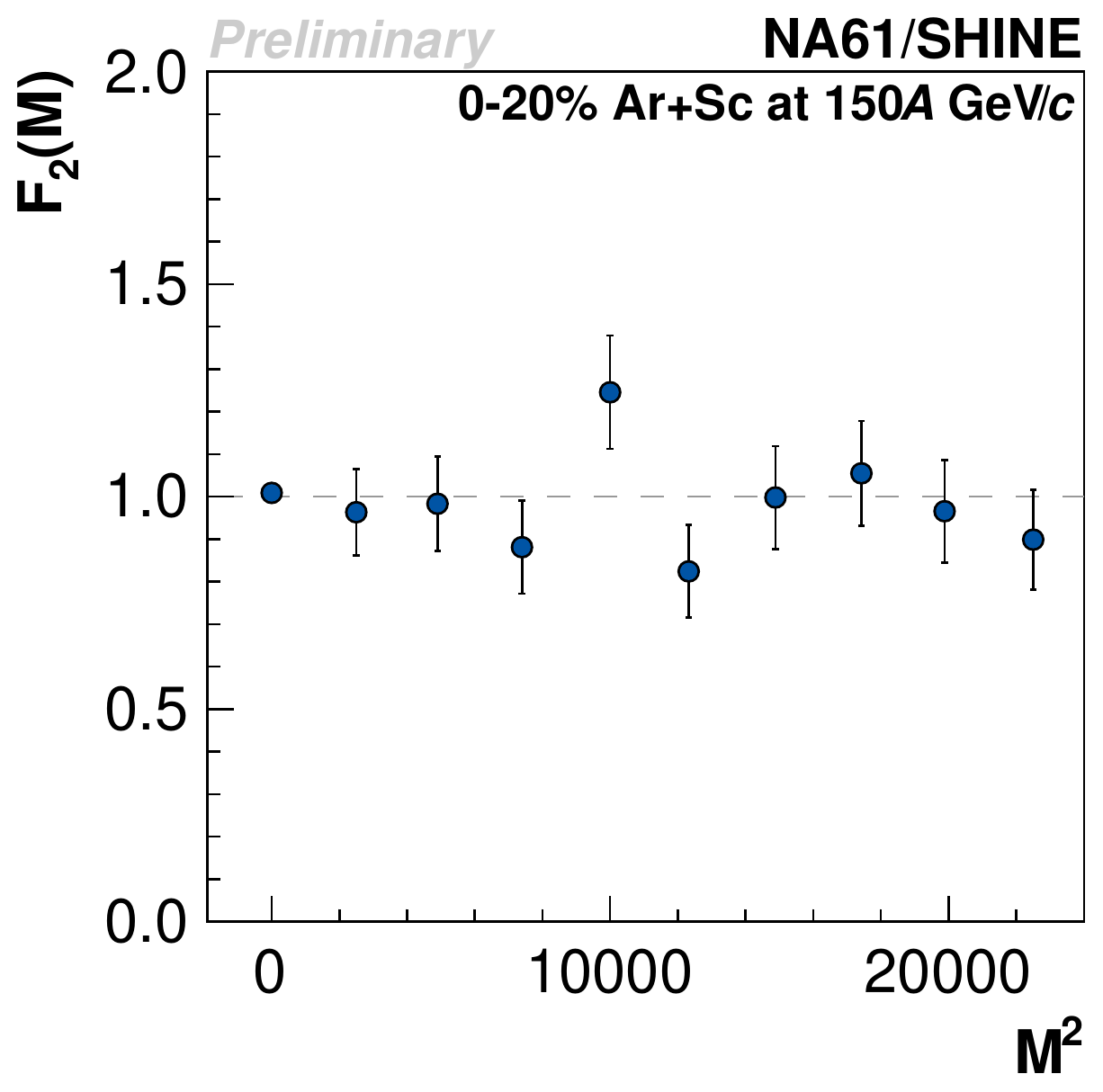}\qquad
    \includegraphics[width=0.42\textwidth]{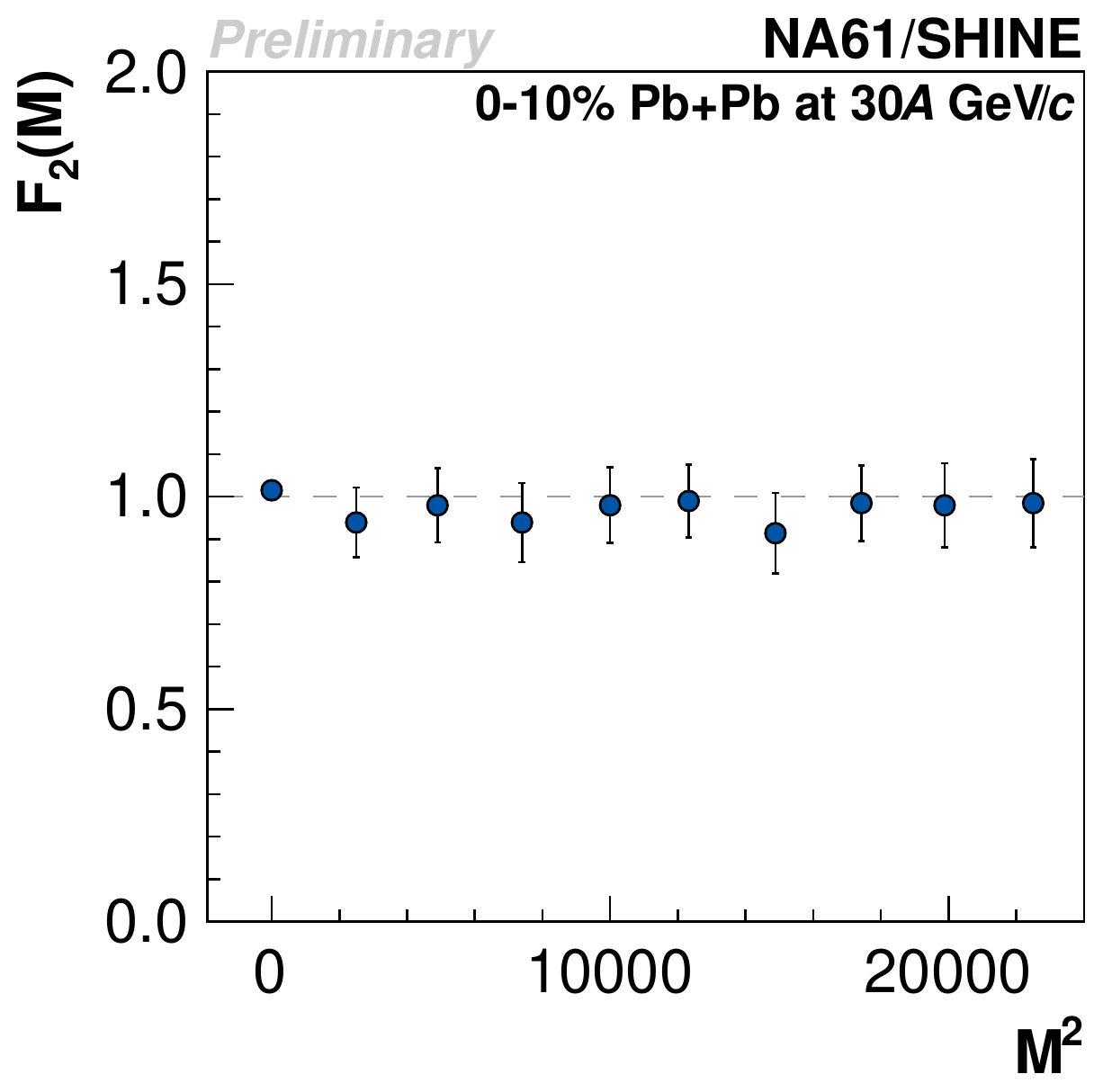}
  }
  \caption{Preliminary results on the midrapidity proton intermittency for
    0--20\% most central \ArSc collisions at \SIa{150}{\GeVc} beam momentum (left) and
    0--10\% most central \PbPb collisions at \SIa{30}{\GeVc} beam momentum (right).}
  \label{fig:p-intermittency}
  \centerline{
    \includegraphics[width=0.42\textwidth]{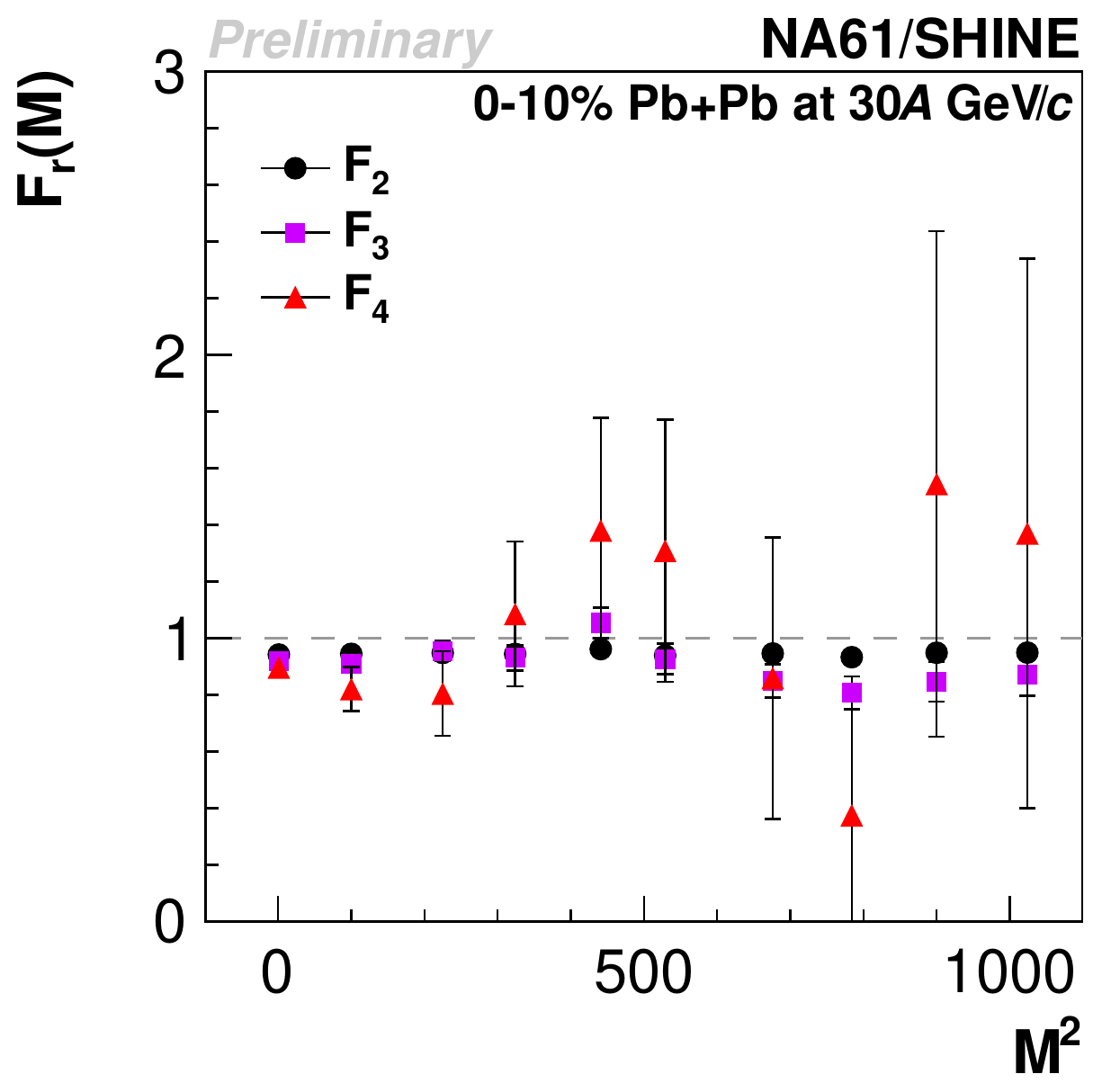}
  }
  \caption{Preliminary results on the midrapidity charged hadron intermittency for
    0--10\% most central \PbPb collisions at \SIa{30}{\GeVc} beam momentum.}
  \label{fig:h-intermittency}
\end{figure}
Another phenomenon frequently associated with the critical point of strongly interacting matter is intermittency.
In
this analysis transverse momentum space in midrapidity is divided into cells in
which particles are counted to calculate scaled factorial moments~$F_r$. For
the system experiencing critical behaviour, the theory predicts~\cite{Antoniou:2006zb} specific
power-law scaling of the moments as a function of the number of cells $M$:
$F_r(M) \sim M^{\phi_r}$. \Figref{fig:p-intermittency} shows preliminary
results on the proton intermittency in central \ArSc and \PbPb collisions at
respectively \SIa{150}{\GeVc} and \SIa{30}{\GeVc} beam momenta.
\Figref{fig:h-intermittency} shows results for charged hadron intermittency in
central \PbPb collisions at \SIa{30}{\GeVc} beam momentum up to the fourth
scaled factorial moment. These analyses feature statistically independent
points and instead of the ($p_x$, $p_y$) cells, cumulative variables are used
to remove dependence of the moments on the momentum
distribution~\cite{Bialas:1990dk}. While no indication of the critical point is
visible in these analyses,
a considerable progress is presently being accomplished in order to provide a more advanced methodology,
more sensitive to intermittency in the presence of sizeable non-critical background~\cite{Nikos}.

\begin{figure}[htb]
  \centerline{
    \includegraphics[width=0.5\textwidth]{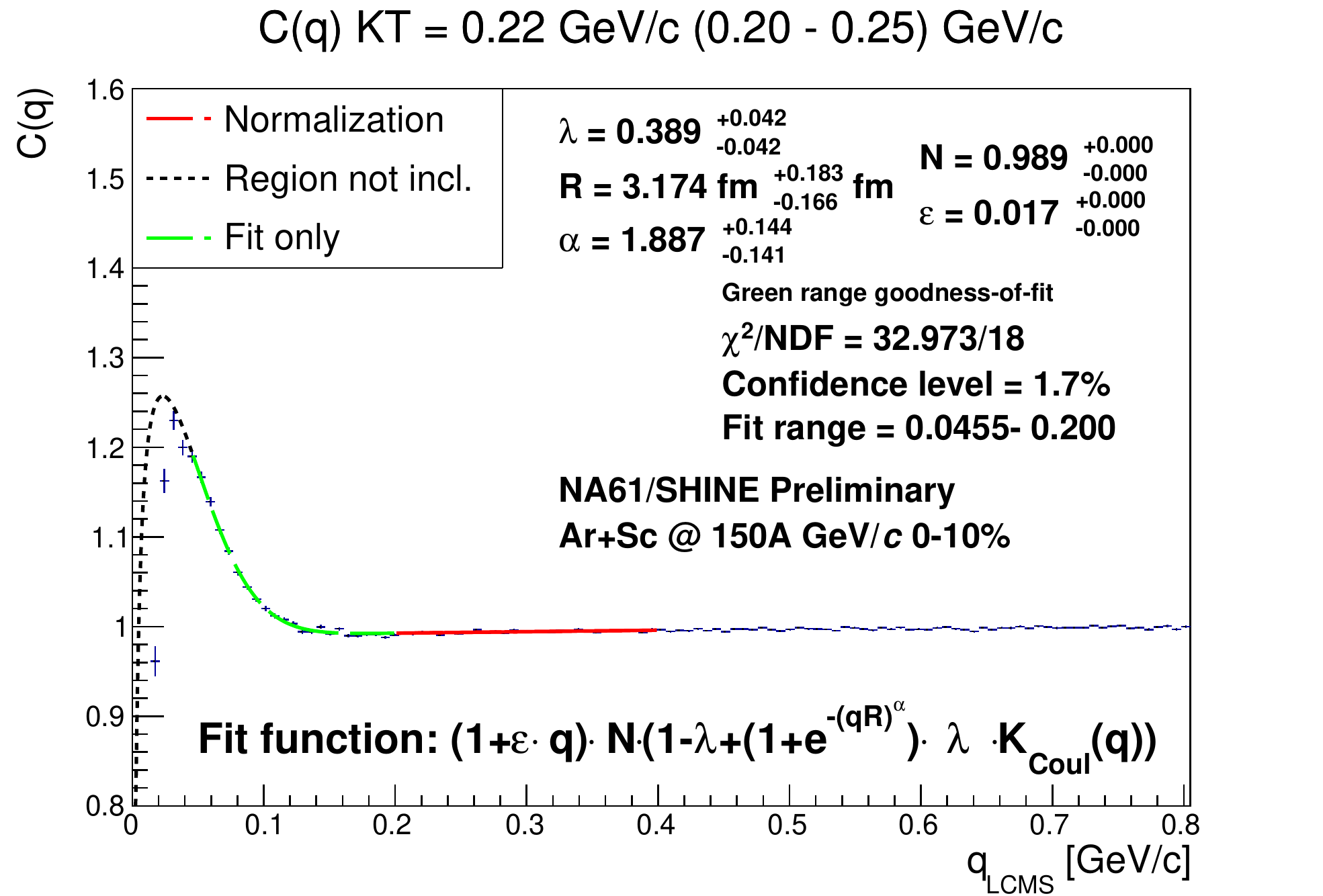}
    \includegraphics[width=0.5\textwidth]{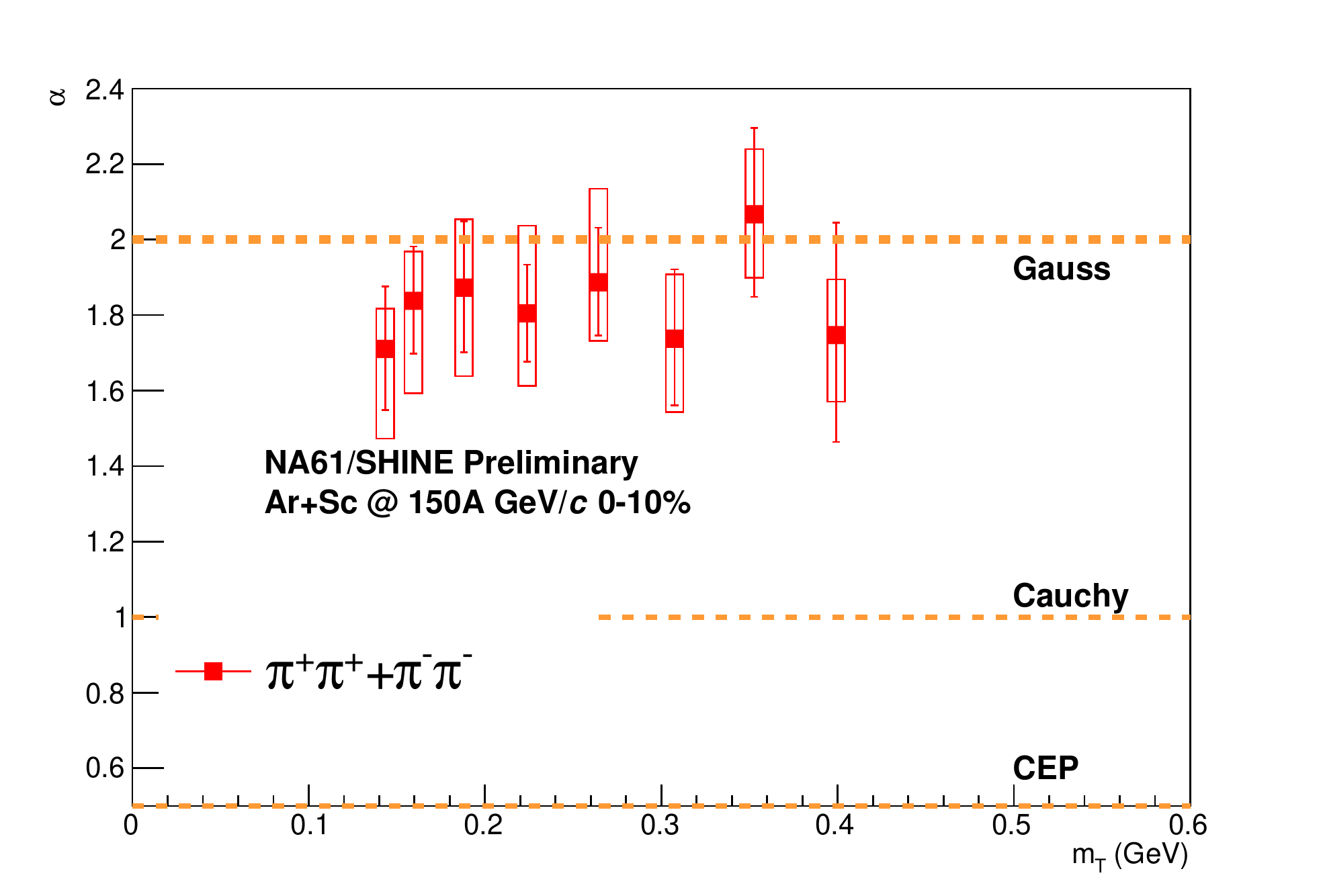}
  }
  \caption{Preliminary results on the symmetric Levy HBT correlations for
    same-charge pion pairs in central \ArSc collisions at \SIa{150}{\GeVc} beam
    momentum. The left plot shows an example fit of the correlation function and
    the right one shows the dependence of the fitted source shape parameter on the
  transverse mass of the pair.}
  \label{fig:hbt}
\end{figure}
The last CP-related results are symmetric Levy HBT correlations for
same-charge pion pairs in central \ArSc collisions at \SIa{150}{\GeVc} beam
momentum shown in \figref{fig:hbt}. Here instead of a usual Gaussian source
shape, a more general Levy-stable distribution is used. Its parameter $\alpha$
describes the shape of the source: for $\alpha = 2$ the source is Gaussian, for
$\alpha = 1$ we have a Cauchy distribution and the 3D Ising model with random
external field predicts $\alpha = 0.5 \pm 0.05$ for a critical system. From the
plot it is clear that there is no indication of the CP for central \ArSc
collisions at the top SPS energy.

\begin{figure}[htb]
  \centerline{\includegraphics[clip, trim=0 0 1018 0, width=0.5\textwidth]{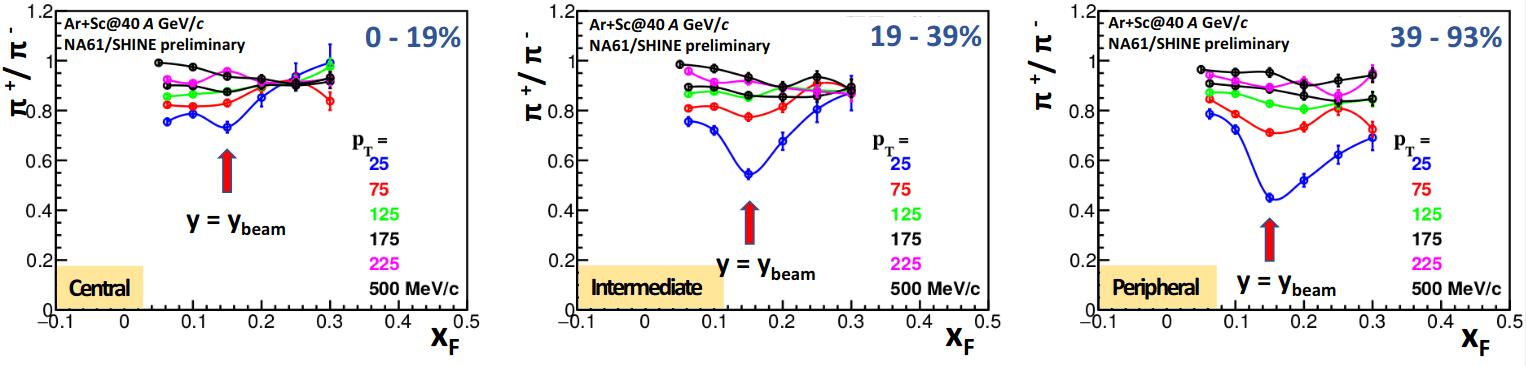}%
  \includegraphics[clip, trim=509 0 509 0, width=0.5\textwidth]{electro.png}}
  \centerline{\includegraphics[clip, trim=1018 0 0 0, width=0.5\textwidth]{electro.png}}
  \caption{Preliminary measurements of the spectator-induced electromagnetic
    effects visible as a depletion of the $\pip/\pim$ ratio for small
    transverse momenta and pion rapidities close to the beam rapidity  for
    central (top, left), intermediate (top, right) and peripheral (bottom)
    \ArSc collisions at \SIa{40}{\GeVc} beam momentum. Top right corners of
  each plot indicate approximate centrality percentiles.}
  \label{fig:em}
  \vspace{1ex}
  \centerline{
    \includegraphics[height=0.22\textheight]{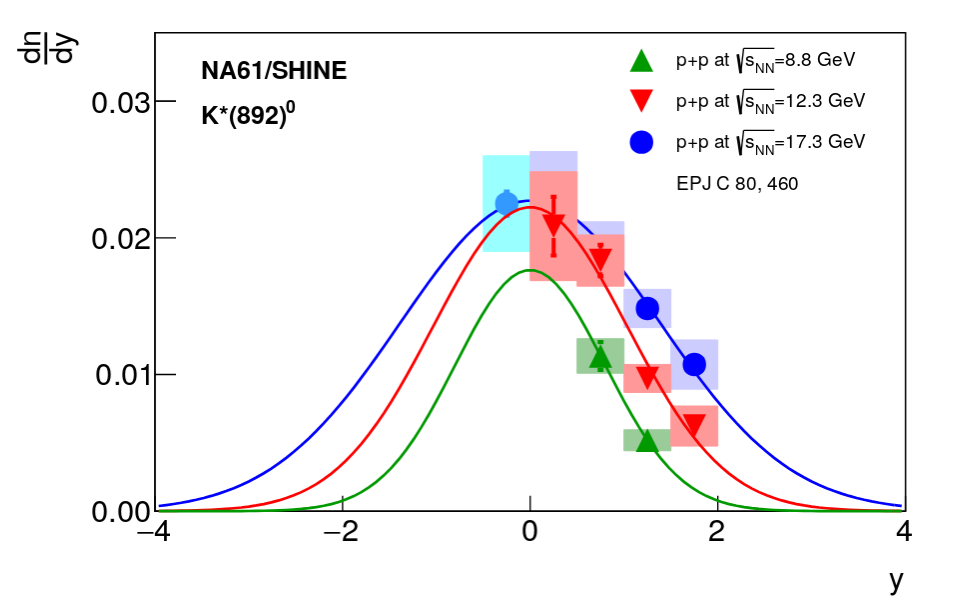}
    \includegraphics[height=0.22\textheight]{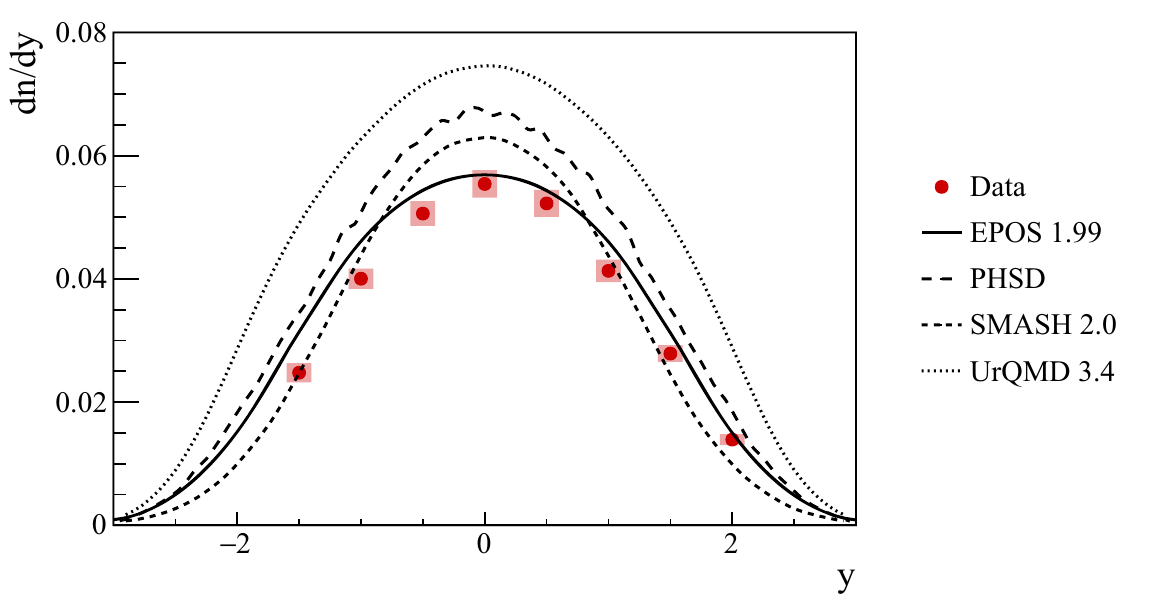}
  }
  \caption[]{New results on rapidity spectra of neutral kaons in \pp collisions:
    the \Kstar at 40 and \SI{80}{\GeVc} beam momenta~\capcite{NA61SHINE:2021wba}
    compared to \SI{158}{\GeVc}~\capcite{Aduszkiewicz:2020msu} (left), the \Kshort at
    \SI{158}{\GeVc} compared to model predictions~\capcite{NA61SHINE:2021iay}
    (right).}
  \label{fig:newpp}
\end{figure}
\section{Spectator-induced electromagnetic effects in \ArSc collisions}
\Figref{fig:em} shows the first time ever observation of spectator-induced
electromagnetic effects in peripheral small systems (\ArSc collisions) at the CERN SPS.
The effect is visible as a depletion of the
$\pip/\pim$ ratio for small transverse momenta and pion rapidities close to the
beam rapidity. Even at intermediate centrality, \ie with a small spectator
system, the effect is strong enough to break isospin symmetry.
Spectator-induced electromagnetic effects provide information on the space-time
evolution of the system~\cite{Ozvenchuk:2019cve}, partially complementary to
the femtoscopy.

\section{New data on hadron spectra in \pp reactions}
The new results on hadron spectra in \pp collisions are also discussed in more
detail in Ref.~\cite{Maciek}. Here only two most recent results on rapidity
spectra of neutral kaons are presented in \figref{fig:newpp}: the \Kstar at 40
and \SI{80}{\GeVc} beam momenta~\cite{NA61SHINE:2021wba} compared to
\SI{158}{\GeVc}~\cite{Aduszkiewicz:2020msu} and the \Kshort at \SI{158}{\GeVc}
compared to model predictions~\cite{NA61SHINE:2021iay}. These data will on the
one hand serve as reference for future data from larger systems and on the
other hand serve as input to models, which struggle to describe the strangeness
production at SPS energies. The latter is visible in the right plot of
\figref{fig:newpp}.

\section{Hardware upgrade and future measurements}
All of the above results were possible thanks to the versatility of the
\NASixtyOne detector system.
The spectrometer
based on large-volume time projection chambers (TPC) features large acceptance
covering the full forward hemisphere down to $p_T=0$. The latter is feasible
due to the fixed-target setup and the magnetic field oriented perpendicular to
the beam direction.
During the CERN Long Shutdown~2, the whole system underwent an upgrade
including construction of a new Vertex Detector, new Beam Position Detectors, new
Time-of-Flight detectors, larger forward calorimeter (Projectile Spectator
Detector), new trigger and data acquisition system and most importantly
replacement of the TPC read-out electronics to increase the data rate tenfold
to 1\,kHz.
\par
The upgrade was necessary to perform the first ever open charm measurement at
the SPS energies, which is the main goal of \NASixtyOne data taking in years 2022--2024 (\figref{fig:programme} left, grey boxes). The latter is
motivated by 3 questions:
  \begin{itemize}
    \item What is the mechanism of open charm production?
    \item How does the onset of deconfinement impact open charm production?
    \item How does the formation of quark-gluon plasma impact $J/\psi$ production?
  \end{itemize}
In order to answer these questions it is necessary to know the full phase space
production of $c\bar{c}$ pairs, for which model predictions vary by 2 orders of
magnitude~\cite{Aduszkiewicz:2309890}.
\par
\textbf{Acknowledgements:}
This work was supported by the Polish Minister of Education and Science (contract No. 2021/WK/10).

\bibliographystyle{h-physrev-acta}
\bibliography{bibliography}

\end{document}